# Ion getter pumps

*C. Maccarrone, P. Manassero, C. Paolini*
Agilent Technologies S.p.A., Via Fratelli Varian 54, 10040 Leinì, Italy

**Abstract**
Ion Getter Pumps (IGP) are used to create ultra-high-vacuum. The IGP operation is triggered by the Penning cell structure, which uses a combination of electrical and magnetic fields to confine electrons and start ionization and getter pumping processes. During the years, more IGP configurations have been developed to cope with different gases, such as reactive gases, hydrogen or noble gases. When the IGP is required to reach pressures lower than $10^{-8}$ mbar, it needs to be baked out with the whole vacuum system, in order to accelerate the release of the gas atoms trapped inside the pump materials, and the leakage current issue has to be taken into account for a reliable pressure reading.

## 1   Introduction

Ion getter pumps (IGP) are devices able to create and maintain ultra-high-vacuum, reaching pressures as low as $10^{-12}\,mbar$. Lower pressures are in principle achievable, but their measurement is particularly challenging and strongly depends on the outgassing of the system on which the ion pump is mounted.

With respect to other vacuum pumps, such as turbomolecular pumps or primary pumps, IGP have some characteristics that make them unique.

First, ion pumps are closed pumps: they do not have any foreline. In fact, an IGP keeps inside all the pumped gas: whatever is pumped by an IGP, will remain in it. This avoids the risk of venting the system to which an IGP is connected, prevents any contamination which could come from the roughing line and does not require a backing pump during operation, but only at start-up since IGP cannot be started at atmospheric pressure.

Furthermore, the ion pump is a static device. It has no moving parts, so it is vibration free and, consequently, it doesn't need any lubricant which could be a source of contamination.

In this paper, we will describe how an Ion Getter Pump works, pointing out the pumping mechanisms for different gases and referring to different pump types. Besides, some relevant topics will be discussed, such as the outgassing process and the leakage current.

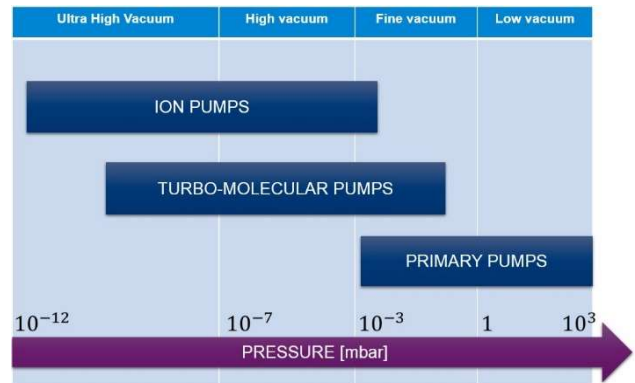

**Fig. 1:** IGP pressure range, compared with turbomolecular and primary pumps

## 2 IGP structure

- **Element** is made up of one or more Penning cells. It is the active part of the pump, by which the gas is pumped. Referring to the standard element named *Diode*, it is made of cylindrical cells (the anode) and two titanium plates (the cathode). The cathode is grounded, while the anode cells are fed with positive high voltage, typically 3 - 5 - 7 kV.
- **Magnets** are positioned outside the pump body to complete the Penning cell structure, creating an axial magnetic field (typically in the range 1200 ÷ 1400 Gauss).
- **Feedthrough** (F/T) is the electrical connector used to feed the pump with high voltage. The central rod of the feedthrough is connected to the element (either to the anode or the cathode, depending on the element type) by means of a metal tab or a wire.
- **Inlet flange** is used to connect the IGP and the vacuum chamber. The inlet flange diameter is one of the parameters determining the conductance, which affects the pumping speed of an IGP.

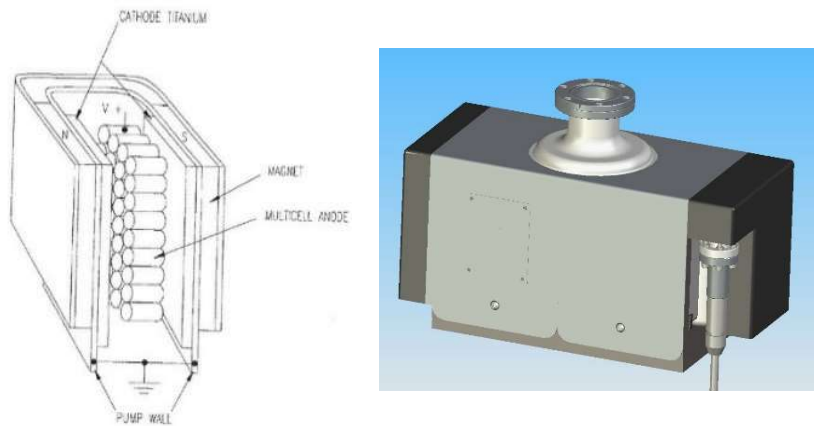

**Fig. 2**: (a) Section of the *Diode* element, where the anode, the cathode and the magnets are indicated; (b) a drawing of an IGP, where the F/T and the inlet flange are indicated.

# 3 Principles of operation

## 3.1 Ionization (of gas molecules)

The operating principle of IGP is based on the Penning cell [1], that is usually made of a stainless-steel anode and two opposite titanium cathodes, which uses a combination of static electric and magnetic fields to achieve confinement of electrons. The electrical field traps the electrons in the axial direction, the magnetic field - in the radial direction.

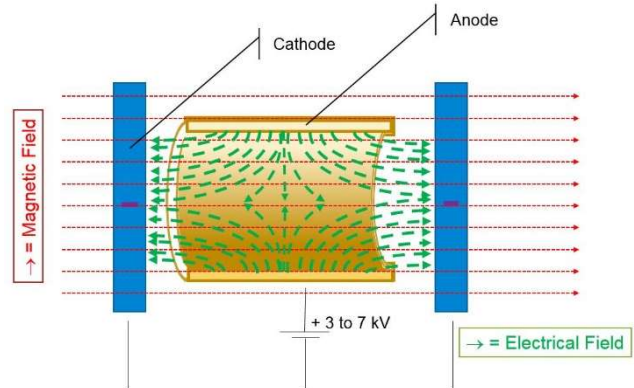

**Fig. 3:** Representation of the Penning cell structure into an IGP.

The aim of the Penning cell is to trap electrons and increase as much as possible the probability to ionize the residual gas, that will be pumped by means of the following mechanism:

– free electrons are produced by applying high voltage between the anode and the cathodes and here are trapped into the Penning cell. This "trapping" increases the path of electrons from cathode to anode by many orders of magnitude and hence increases the probability of a collision between electrons and gas molecules;

– because of collisions, the trapped electrons ionize the background gas, thus creating positive ions and extracted electrons;

– ions, positively charged, are attracted towards the cathodes;

– extracted electrons are trapped as well.

This mechanism can be considered the starting point for IGP operation and occurs in every IGP independently from the element type. However, the pumping mechanism in its entirety changes in the presence of different gases, since the ionization process has different consequences depending on the properties of the gas atoms.

## 3.2 Pumping of chemically active (*getterable*) gases

*Getterable* gases, such as nitrogen ($N_2$), oxygen ($O_2$), carbon monoxide ($CO$) or carbon dioxide ($CO_2$), are those gases which form chemical compounds with the surface of *getter* materials, such as titanium (the material the cathode is made of).

When gas atoms are ionized and electrically attracted by cathodes, some are neutralized and buried into the titanium cathodes. This mechanism, used by IGP to pump the gas, is called *pumping at the cathode or ionization pumping*.

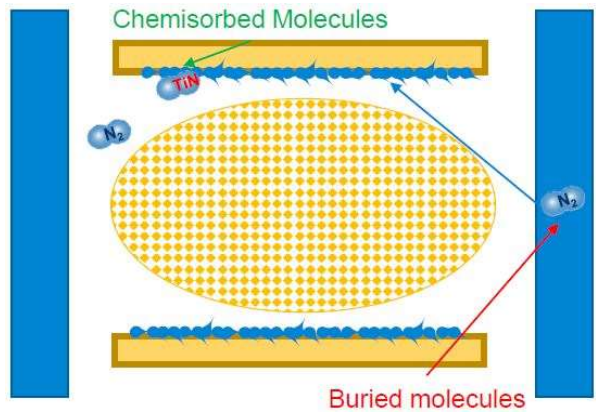

**Fig. 4:** Representation of the pumping mechanism for *getterable* gases

The ions, hitting the cathodes, cause the *sputtering* of the titanium atoms: they are emitted from the cathodes and cover the anode surfaces, creating an active titanium film which chemically traps the background gas molecules. This mechanism is named *pumping at the anode or getter pumping*.

### 3.2.1    *Saturation of an IGP*

While *pumping at the anode* mechanism is permanent, since it works forming chemical bonds, *pumping at the cathode* is not stable: as the sputtering of the titanium atoms from the cathodes goes on, the cathodes are eroded and the previously implanted atoms can be released.

So, the net pumping speed decreases until an equilibrium condition between ion implantation and gas re-emission is reached. At equilibrium, the IGP is called *saturated*. This term is sometimes misleading, since it can be confused with the end of the pump life.

Hence, when starting to use an IGP, its pumping speed will be higher than the nominal one, because the pump is still unsaturated and the pumping at the cathode mechanism is still actively working. The term *saturated* indicates the stable operation mode of an IGP, in which it reaches its nominal pumping speed, and not the end of the pump life. The time the pump needs to reach the equilibrium depends on the pressure: the higher is the pressure, the faster the equilibrium is reached (see Fig. 5).

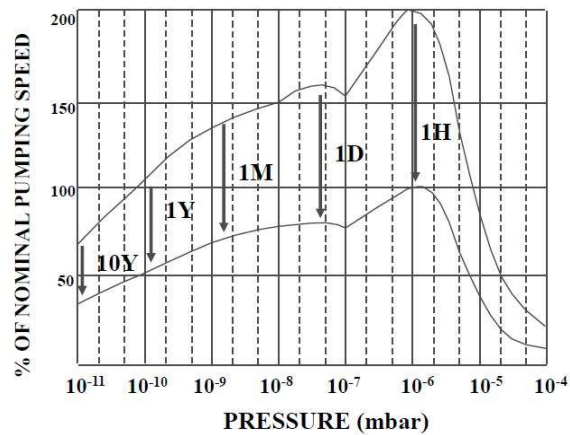

**Fig. 5**: Time needed to *saturate* an IGP depends on the pressure

The cathode erosion occurs not uniformly on the cathode plates, since the electric field lines carry most of the ions in front of the anode cells centre, as shown in Fig. 6

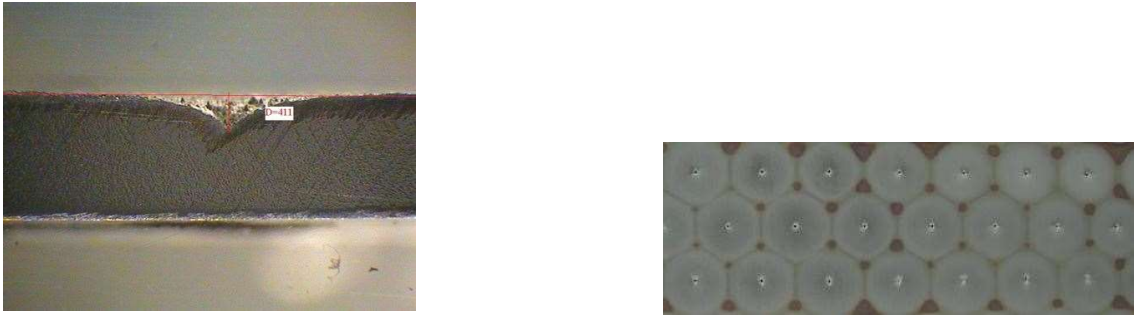

**Fig. 6:** Titanium erosion under microscope (a) and a cathode plate after IGP long term use: the erosion drills the cathode plate in correspondence of the centre of the anode cells (b)

### 3.3 Pumping of hydrogen

Since hydrogen is chemically reactive, it can be chemisorbed by the sputtered titanium film (pumping at the anode).

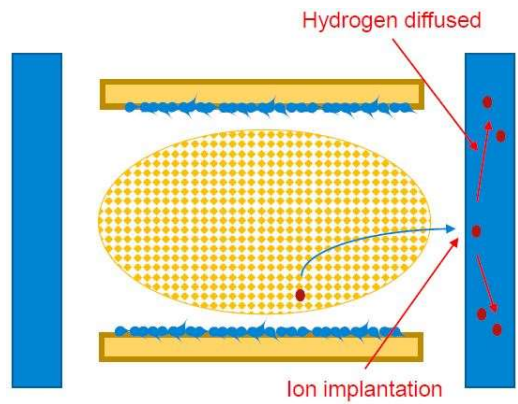

**Fig. 7:** Representation of the pumping mechanism for hydrogen

Besides, hydrogen has a high solubility in titanium, so it diffuses in the cathode after implantation and forms a solid solution with it. This improves the pumping at the cathode mechanism and leads a decrease of the titanium sputtering yield; the erosion of the cathode is negligible and only a small quantity of hydrogen is re-emitted into the pump. As a consequence, the pump remains almost unsaturated for hydrogen and the pumping speed is roughly 50% to 100% higher than for nitrogen.

### 3.4 Pumping of noble gases

Noble gases are not chemically active, so they are not chemisorbed by the titanium film which covers the anode.

They can be implanted in the cathode, but this pumping mechanism is not stable: after saturation, noble gases are released back into the pump and a phenomenon usually called *noble gas instability* occurs. Because of the gas release, the pressure increases of one or few decades. Then the gas is re-implanted in the cathode and the pressure decreases without need of intervention on the system; the situation is stable until the cathode erosion reaches again the implanted gas and another re-emission happens, leading to another pressure increase.

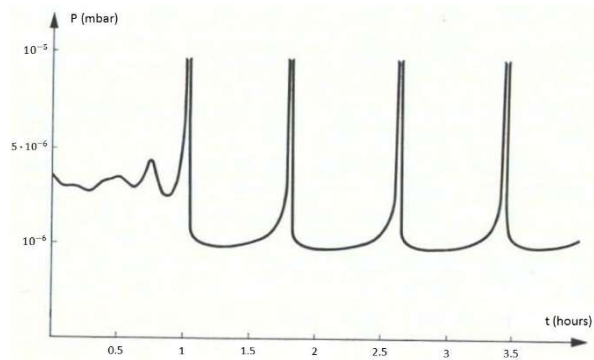

**Fig. 8:** An example of *noble gas instability*

Therefore, a key indicator of *noble gas instability* occurrence is the periodicity of pressure peaks, and the time needed to see such a behaviour depends on the pressure (the amount of pumped gas).

The most stable and efficient mechanism to pump noble gases occurs when gas ions are reflected back from the cathodes as high energy neutrals. Hence, noble gases can be implanted (not chemisorbed) permanently into the titanium that covers the anode. This titanium is not subject to sputtering phenomena; the noble gases buried into it will therefore not be released again in the system.

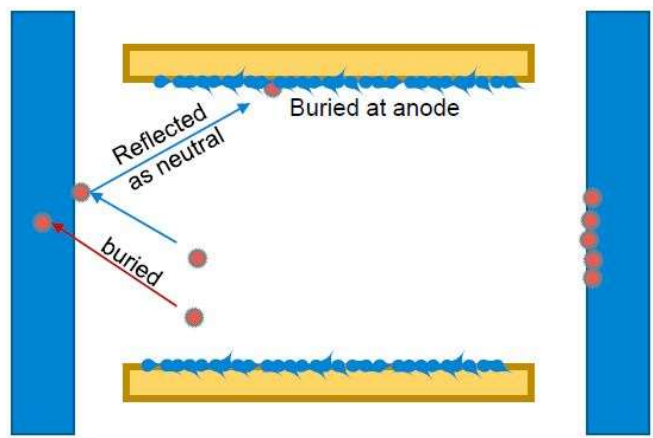

**Fig. 9:** Representation of the pumping mechanism for noble gases

# 4    Agilent IGP elements

As discussed above, the IGP operating principle depends on the gas type. For this reason, different types of IGP were developed to optimize the pump performances when pumping a particular gas type.

So, an IGP optimized for noble gases can be different from an IGP optimized for *getterable* gases.

In particular, ion pump types differ in their geometry, the shape of the element and the materials used for the cathode.

## 4.1 Noble Diode

As discussed in Sec. 2.4, it is required to increase the probability of reflecting ions as high energy neutrals to better pump noble gases. This goal can be achieved by choosing a different material for the cathodes, with a higher atomic number, to increase the probability that this neutralization and reflection occur for the noble gas ions. For this reason, *Noble Diode* IGPs have been developed. Their elements contain one cathode made of tantalum (atomic number Z = 73) instead of titanium (Z = 22).

This solution increases both pumping speed and stability when pumping noble gases with respect to a standard *Diode* pump. In fact, by increasing the probability of reflecting ions as neutrals, the most efficient pumping mechanism is enhanced and *noble gas instability* is prevented. Otherwise, the modified *Noble Diode* element leads to lower performances for hydrogen with respect to a *Diode* IGP, since the solubility in tantalum is lower than in titanium.

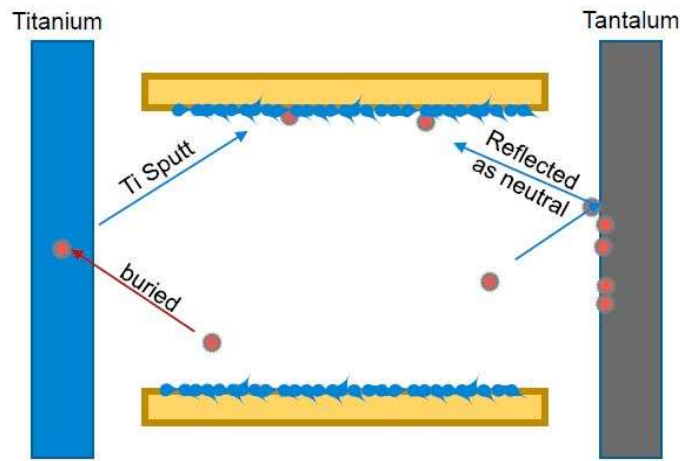

**Fig. 10:** Schematic representation of the *Noble Diode* element.

## 4.2 Triode and *StarCell®*

While the *Noble Diode* element is conceived to better pump noble gases by changing the cathode material, another approach is to change its geometry. Thus, the *Triode* IGP has been developed. Its element has a cathode made of strips that are separated from the pump body.

For this element configuration, negative voltage is applied to the cathode while the anode is grounded and the Penning cell ionization mechanism is still the same as in *Diode* ion pumps.

Because of the *Triode* element geometry the probability to pump the gas at the cathode, that leads to instability for noble gases (see Sec. 2.4), is greatly reduced. On the contrary, ions hitting the cathode with a glancing angle have an increased probability to be emitted as neutrals and, then, to be implanted in the titanium sputtered on the anode and the pump walls, which become active components in pumping the gas.

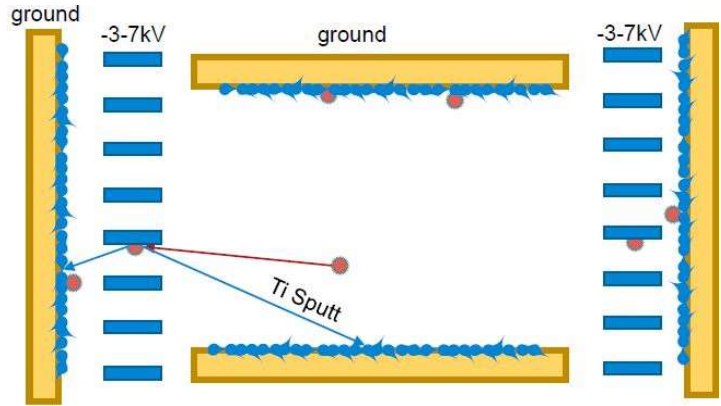
**Fig. 11:** Schematic representation of the *Triode* element.

An improved version of the *Triode* element is the *StarCell* element [2], shown in Fig. 12, conceived and developed by Agilent (formerly Varian) [1].

The shape of the cathode, which is made of stars with small bended wings, is optimized to maximize the reflection of neutrals and to ensure a longer lifetime with respect to a *Diode* pump, since titanium erosion is not centred in a few points (see Par. 2.2.1) but it is distributed more evenly.

With respect to the *Noble Diode* IGP, the *StarCell* IGP better pumps hydrogen since both cathodes are made of titanium.

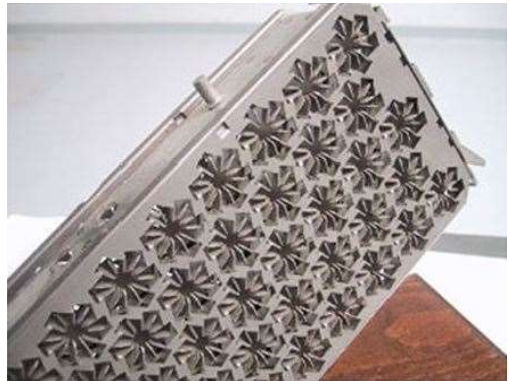
**Fig. 12:** Photo of the *StarCell* element

**Table 1:** Comparison of different Agilent IGPs performances for different gases

| Gas | StarCell [a] | Diode [a] | Noble Diode [a] |
|---|---|---|---|
| Hydrogen | 2 | 3 | 1 |
| Argon (or Helium) | 4 | 1 | 3 |
| Water | 2 | 3 | 2 |
| Methane | 3 | 2 | 3 |
| Nitrogen | 2 | 3 | 3 |
| $O_2$, CO, $CO_2$ | 2 | 3 | 3 |

[a] 1: poor;   2: good;   3: excellent;   4: outstanding

# 5 Lifetime and leakage current

The principal causes of the end of an IGP life are summarized in the following:

1) the erosion completely drills the cathode along its whole thickness, so the sputtering is no more effective and can't refresh the titanium film on the anode. This event is a direct result of the IGP operating mechanism (discussed in Sec. 2).

   The time needed to wear out the titanium depends on the cathode thickness and the working pressure. It can be assumed that ten times higher is the pressure, ten times higher is the sputtering rate and so ten times shorter is the IGP life.

   For Agilent IGP, a rough evaluation of the operating life of a pump working at $1 \cdot 10^{-6}$ mbar (nitrogen) is:

   - 50000 hours for *Diode* and *Noble Diode*
   - 80000 hours for *StarCell* pumps

2) A vacuum leak: if this does not happen at the beginning of the pump life, it is very unlikely to occur on a long-term time scale.

3) Low magnetic field: if the magnetic field is not strong enough, the Penning cell is not able to trap electrons to ionize gas atoms. It may occur if the magnets are overheated (e.g. during the bakeout) and reach a temperature higher than their corresponding Curie temperature.

4) *Noble gas instability*: see Sec. 2.4. This phenomenon may prevent an efficient use of the ion pump on the system.

5) Short circuit: it may occur in the case of complete metallization of the ceramics, which work as electrical insulators through the anode and the parts of the pump at ground potential, or in the case of a remarkable cathode deformation, that can shorten its distance from the anode. The latter, for example, can be due to the absorption of a huge amount of hydrogen.

6) The leakage current, as discussed in the following section. Again, as in the case of the noble gas instability, the IGP will continue pumping even if affected by a leakage current, but the leakage current can make it less efficient as a pressure indicator and, in the most severe cases, can cause the overheating of the pump and consequently increase its outgassing.

## 5.1 Leakage current and SEM element

Ion pumps are often used as pressure indicators, thanks to the well-known almost linear dependence of pressure on the ion current, as long as the cell geometry, the number of cells, the voltage and the magnetic field are taken as fixed parameters [3]. Unfortunately, some spurious currents, independent of pressure, can arise in the ion pump; at low pressures, such currents can be comparable to or much higher than the pressure-dependent ion current.

The leakage current does not actually affect the pumping efficiency at all, but when this phenomenon occurs, the ion pump becomes useless in pressure reading; in the most dramatic case, pressures below $10^{-7}$ mbar cannot be read.

The spurious current can arise from both external sources, such as power supply, the connecting cable or the high-voltage feedthrough, and internal sources. Among internal sources, the metallization of ceramic insulators during operation is potentially responsible for this phenomenon.

However, the main cause of leakage current occurrence is the Field Electron Emission (FEE) from cathodes, by which free electrons are emitted from the titanium surface when strong electric field is applied.

Therefore, the leakage current can limit the use of an ion pump as a pressure indicator, especially in low pressure applications (lower than $10^{-9}$ mbar). In this pressure range, the pumping current, proportional to the pressure, is comparable to or, in some cases, can be even lower than the leakage current due to FEE.

In order to reduce leakage current, a patented [4] anode structure was developed by Agilent, in which interstitial cavities between contiguous cells were eliminated, as shown in Fig 13.

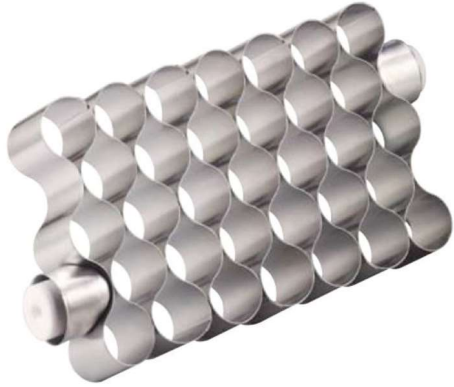

**Fig. 13:** Patented SEM element

Studies have shown that a greater number of dendrites (whiskers), which are sources of field emission, are created in the interstitial areas between the cells. This optimized element is usually called "SEM element" since it is very often employed in scanning electron microscope applications, where the ion pump current is used as a pressure set point to protect the microscope gun (filament or microtip electron emitters) in case of overpressure. It allows a reliable pressure indication, converted from the ion pump current reading, down to the UHV pressure range.

# 6 Outgassing

In solids, about 1 atom on 1000 is a gas atom. Part of this gas is trapped in the atomic structure of the matter, while the other part is on the surface. This last amount of gas, which has been adsorbed at atmospheric pressure, can be released by desorption process when the surface is exposed to vacuum.

Hence, the outgassing is an infinite source of gas which limits the base pressure achievable by a vacuum system. The first way to cope with outgassing is to choose low-outgassing materials, but this solution is not sufficient to reach ultra-high-vacuum level. Since we know that outgassing quickly increases with temperature and slowly decreases in time, vacuum systems and vacuum components in general, and ion pumps among them, are usually baked at ∼250°C, in order to accelerate the release of adsorbed and trapped gas atoms. It is worth noticing that the bakeout can be effective only if performed on all the vacuum system components. In fact, it is of course not sufficient to bake only an IGP.

As shown in Fig. 14, during the bakeout the pressure increases as a consequence of the enhanced outgassing process, but once the system and the pump are cooled to room temperature, an improvement of the reached base pressure is observed with respect to the pressure that would be achieved by the system in the same time without the bakeout.

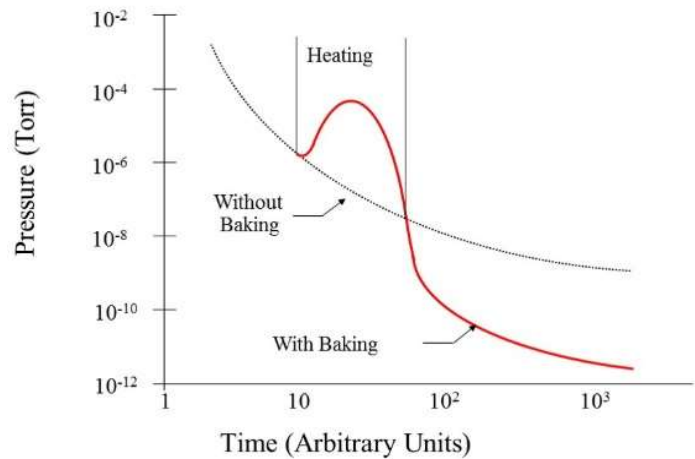

**Fig. 14:** Plot representing the pressure as a function of time when the pump is backed out

Another way to limit the outgassing rate during the pump operation is to provide preliminary thermal treatments to the pump components. In particular, the *Vacuum Firing* process, during which the material is baked at high temperature (e.g. 950 °C in the case of stainless steel) in a vacuum furnace, is used to eliminate hydrogen atoms from cathode and anode surfaces (in some cases from the whole pump body) before IGP is assembled and finally baked out.

For Agilent VIP 200, which was launched in the market in 2016, the *Vacuum Firing* process has been applied to all the surfaces exposed to vacuum, in order to further reduce the hydrogen outgassing rate and to accelerate the pumpdown to ultimate pressure.

## 7    Conclusions

Since their invention in 1957, ion pumps have continuously moved towards lower pressures and are still today the technology of choice for most ultra-high vacuum (UHV) applications, in both research and industry. This includes a large variety of vacuum systems including particle accelerators, synchrotron light sources, gravitational wave detectors, electron microscopes, surface analysis and medical equipment.